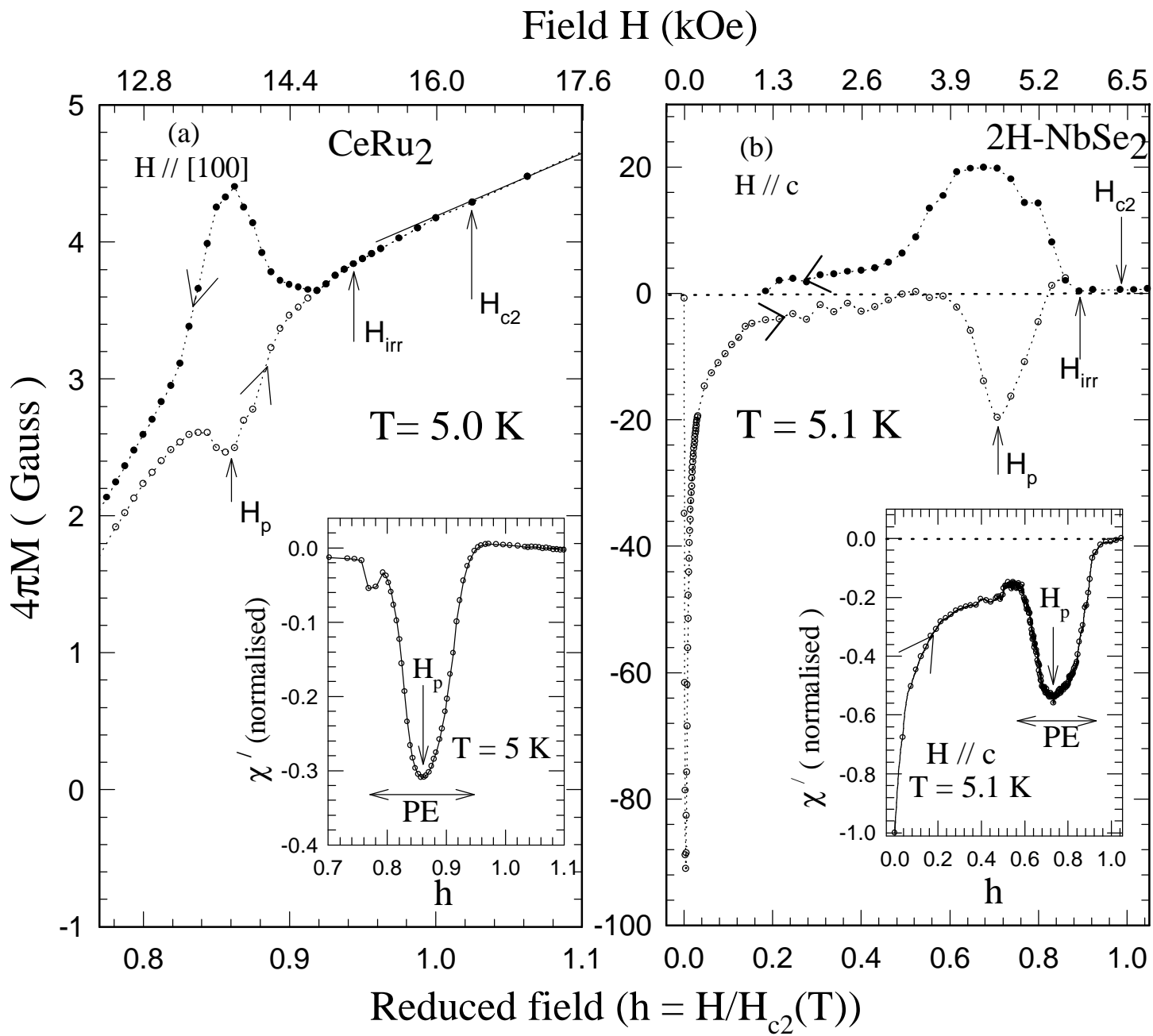

Fig. 1 (S. S. Banerjee et al)

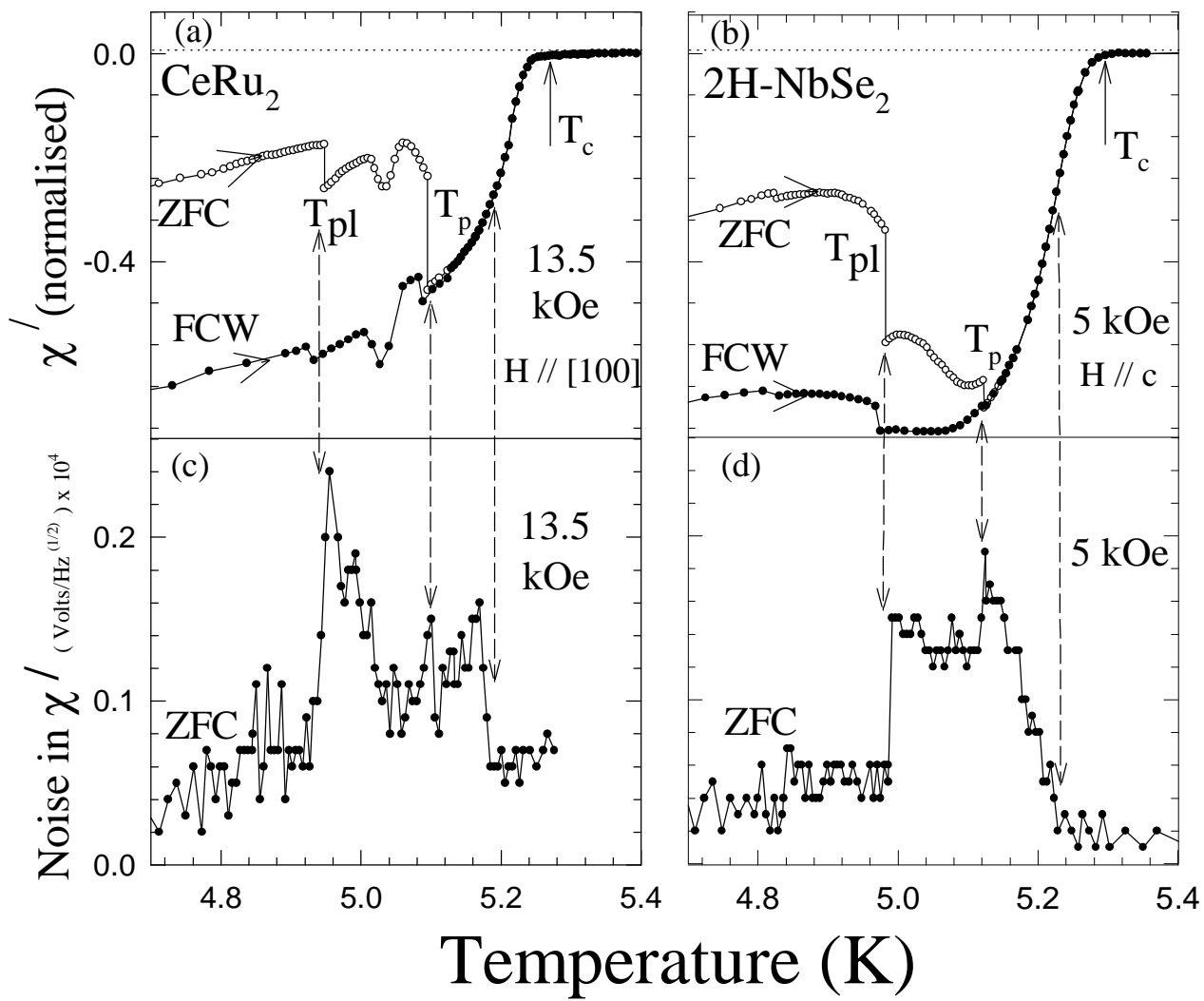

Fig. 2 (S. S. Banerjee et al)

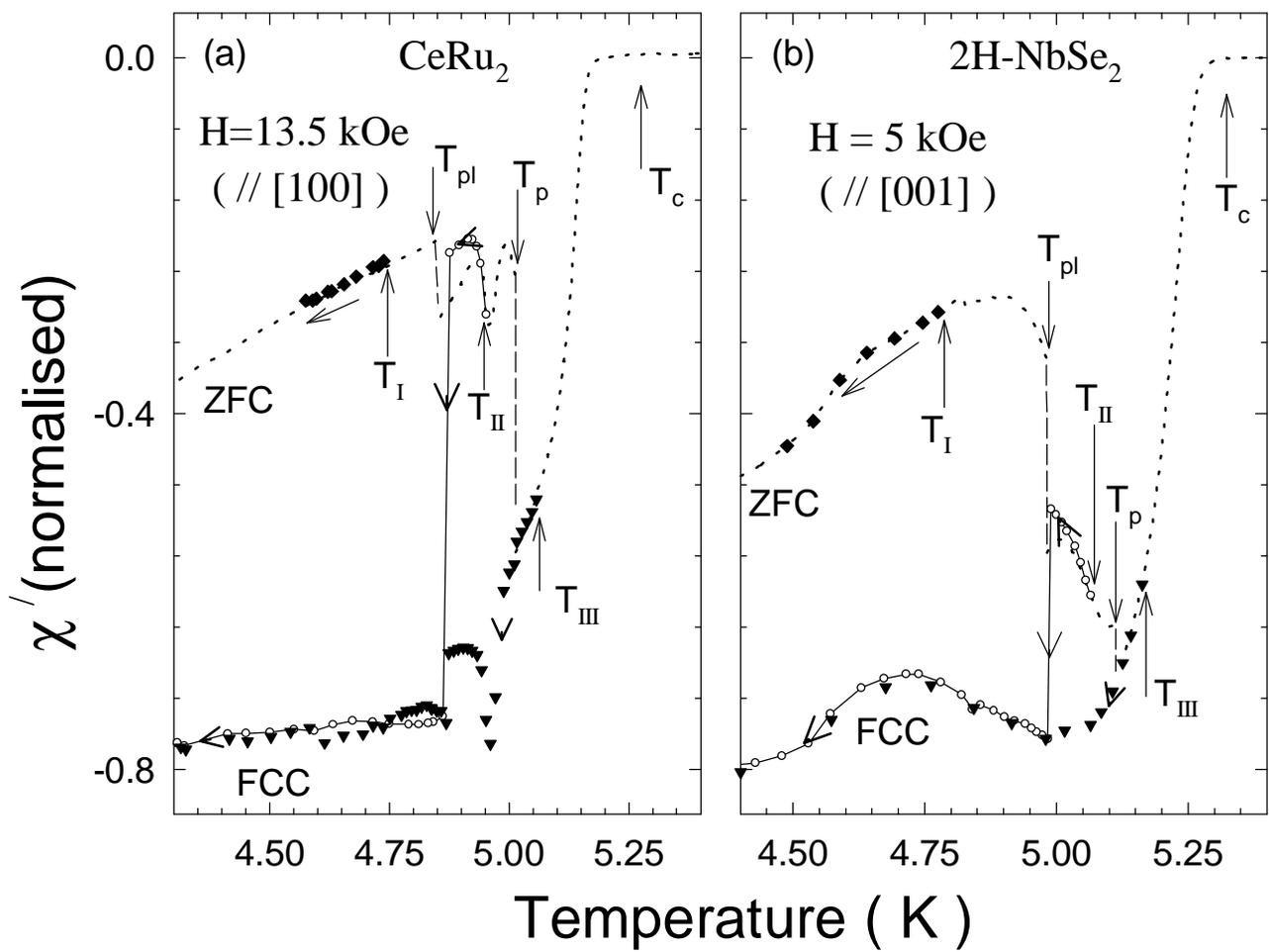

Fig. 3 (S. S. Banerjee et al)

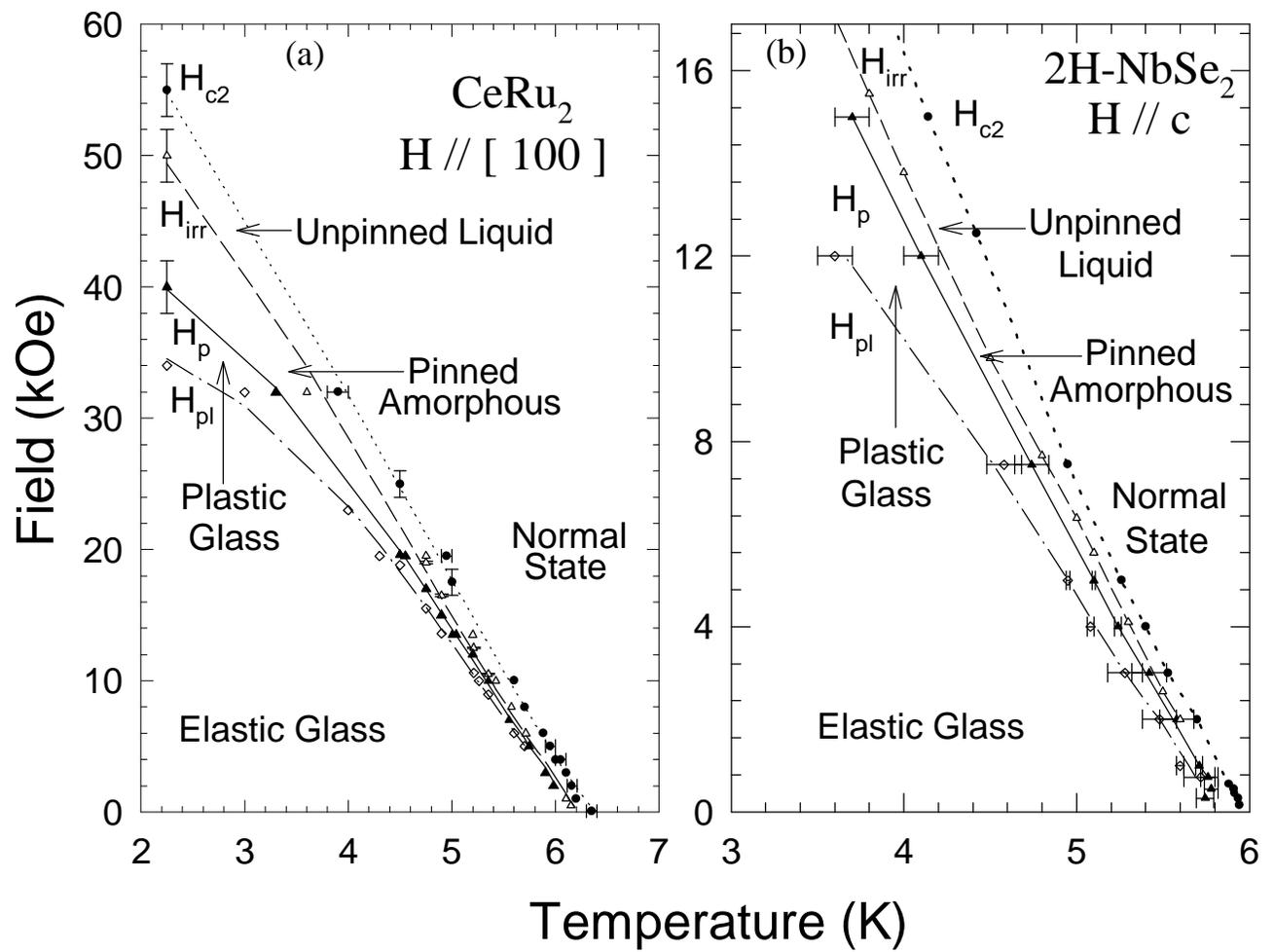

Fig. 4 (S. S. Banerjee et al)

# Anomalous Peak Effect in CeRu$_2$ and NbSe$_2$ : Fracturing of a Flux Line Lattice ?

S. S. Banerjee[1], N. G. Patil[1], Subir Saha[1], S. Ramakrishnan[1], A.K. Grover[1,*],
S. Bhattacharya[1,3,*], G. Ravikumar[2], P.K. Mishra[2], T.V. Chandrasekhar Rao[2],
V.C. Sahni[2], M. J. Higgins[3], E. Yamamoto[4], Y. Haga[4], M. Hedo[5], Y. Inada[5] and Y. Onuki[5]

[1]Tata Institute of Fundamental Research, Mumbai-400005, India
[2]TPPED, Bhabha Atomic Research Centre, Mumbai-400085, India
[3]NEC Research Institute, 4 Independence Way, Princeton, New Jersey 08540
[4]Advanced Science Research Centre, Japan Atomic Energy Research Institute, Tokai, Ibaraki 319-11, Japan
[5]Faculty of Science, Osaka University, Toyonaka 560, Japan

CeRu$_2$ and 2H-NbSe$_2$ display remarkable similarities in their magnetic response, reflecting the manner in which the weakly pinned flux line lattice (FLL) loses spatial order in the Peak Effect (PE) regime. The discontinuous change in screening response near the onset of PE and the history dependence in it are attributed to a disorder-induced fracturing transition of the FLL, as an alternative to the scenario involving the appearance of a spatial modulation in superconducting order parameter in CeRu$_2$.

A sharp transition between weak and strong irreversibility in the mixed phase of the cubic Laves (C15) superconductor CeRu$_2$ has received a great deal of attention recently [1–6]. Magnetization studies of this phenomenon reveal similarities with the peak effect (PE) [7] ( an anomalous peak in the critical current density $J_c$ below $H_{c2}$), also a subject of current interest in a variety of other systems [8–13]. Modler et al [3] have shown a magnetic phase diagram in CeRu$_2$ similar to that for the antiferromagnetic heavy fermion superconductor UPd$_2$Al$_3$ and attributed the *transition to strong irreversibility* to the appearance of a generalized Fulde-Ferrel-Larkin-Ovchinnikov (GFFLO) state, characterized by a spatially modulated superconducting order parameter [4,14]. It is believed to occur in systems with large normal state electronic susceptibility [4]. For the GFFLO state, vortex line(s) can get segmented into coupled pieces of lengths comparable to wavelength of modulation in order parameter and these segments could conform more easily to the inevitably present pinning centres thereby enhancing pinning. However, a clear theoretical support [14] for such a scenario in the observed parameter space , i.e., at T >> 0.56 $T_c(0)$ and H ≈ 0.1 $H_{c2}(0)$ [6], is lacking.

Amongst other superconductors showing PE, the hexagonal 2H-NbSe$_2$ is being widely pursued [9–13] not only due to its weak pinning property but also due to pronounced nature of PE in it, whose details appear related to flux line lattice (FLL) melting [9]. In this system, a variation in quenched disorder produces a rich evolution of the PE phenomenon with characteristic new features [12]. Indeed, recent transport $J_c$ measurements along different (H,T) paths in CeRu$_2$ [5] show similarities with those in 2H-NbSe$_2$ [10] in the PE phenomenon. In latter compound, the normal state susceptibility is small [9] and it is not a candidate for the GFFLO state.

The above facts raise the question : Is the PE in CeRu$_2$ of an origin different from that in NbSe$_2$? We present here the results of dc and ac magnetization studies in both systems, which show their remarkable similarity. In particular, a discontinuous transition occurs in the in-phase ac susceptibility ($\chi'$) near the onset temperature of PE in both systems. We propose that an explanation of the observed phenomenon is to be found in the process of *loss of order* of the FLL , namely, the occurrence of a fracturing transition of the FLL and does not, a priori, require the GFFLO state in CeRu$_2$ [3,4].

The experimental data are obtained from a standard Quantum Design Inc. SQUID magnetometer and an ac susceptometer [15]. The crystal of CeRu$_2$ (2.75 x 1.45 x 0.9 mm$^3$) with $T_c(0) \approx 6.3$ K is the same as the one used for a de-Haas van Alphen study [16] and the crystal of 2H-NbSe$_2$ (2 x 1.5 x 0.3 mm$^3$) with $T_c(0) \approx 6$ K is from the same batch as utilised by Henderson et al [10]. In the two crystals, the values of $R_{300K}/R_{7K}$ are 13 and 11, respectively.

The main panels of Fig.1 show the dc magnetisation hysteresis data in the PE region [17] in both the crystals at (about) 5 K. For 2H-NbSe$_2$, the complete magnetisation cycle is displayed in Fig. 1(b), whereas in CeRu$_2$, only the anomalous opening and subsequent closing of the hysteresis bubble in the PE region has been shown in Fig. 1(a). The field at which bubble is widest identifies the peak field H$_p$ and the collapse of hysteresis locates the irreversibility field H$_{irr}$, above which the differential magnetisation $\Delta$M/$\Delta$H is positive [18]. The upper critical field H$_{c2}$ is marked by the onset of diamagnetic contribution due to superconductivity and is determined as the limiting field (from the high H end) down to which (normal state) paramagnetic response displays the linear variation with H (see Fig.1(a)). The inset panels in Figs.1(a) and 1(b) show H dependence of $\chi'$ at about 5 K. The PE is accompanied by a rapid enhancement in the diamagnetic $\chi'$ response, implying enhanced pinning.

Earlier studies [3] have shown that the onset of PE is discontinuous and hysteretic, characteristic of a first order transition. The abrupt onset of PE in our measurements is seen clearly in $\chi'$(T), in the upper panels of Fig.2, where we show data for two magnetic histories. In one, the system is first cooled to the lowest T in zero-field (ZFC) and then the field is applied, and in the other, it is cooled in a field (FC) to the same temperature. $\chi'$ data are recorded during warm up (W). Note that the difference in $\chi'$ between ZFC and FCW curves occurs over a wide temperature range. This difference implies the importance of disorder in the FLL as it reaches different (metastable) states for the same H, T values. (This is reminiscent of magnetic response in spin glasses [19]). In a simple description (Bean's Critical State Model [20]), $\chi' \approx$ (-1 + $\alpha$ h$_{ac}$ / J$_c$), where $\alpha$ is a size and geometry dependent constant and h$_{ac}$ is the ac field. The FC state produces higher diamagnetic screening response, implying the FLL is more strongly pinned (i.e., more disordered) with a larger J$_c$ than that in the ZFC state [10]. Within the Larkin-Ovchinnikov [21] description of collective pinning, the pinning force is given by, F$_p$ = J$_c$B = (n$_p$<f$_p^2$>/V$_c$)$^{1/2}$, where f$_p$ is the elementary pinning force and n$_p$ is the density of pins [22] ; we conclude that in ZFC case, FLL is much more correlated with larger correlation volume V$_c$.

The onset of PE in ZFC branch in both cases ( cf. Figs. 2(a) and 2(b) ) manifests as a resolution-limited jump of width < 1mK (all data not being shown) in $\chi'$ at T = T$_{pl}$. Above T$_{pl}$, $\chi'$ shows considerable structure, notably, a smooth local minimum followed by yet another jump at T$_p$. Somewhat above T$_p$ , most of the difference between ZFC and FCW values disappears. The more disordered FC state is understood to arise from the supercooling ( from above T$_p$ ) of the FLL state with frozen-in liquid-like correlations of the flux assembly. *We infer that the FC branch represents much smaller regions of correlated lattice as compared to those in the ZFC branch, consistent with earlier neutron scattering experiments [2] in CeRu$_2$.*

We attribute the sharp anomalies in $\chi'$ at T$_{pl}$ and T$_p$ to be distinct first order transitions, involving sharp decreases in the correlation volume V$_c$ of the FLL. Further studies [23] confirm that these transitions are strongly affected by disorder, since their magnitude can be controlled, for example, by the impurity level [12], thereby altering the effective pinning.

We propose that near the onset of the PE, the first order transition at T$_{pl}$ is caused by a fracturing of the FLL, which leads to a plastically deformed lattice, that suffers further loss of order at T$_p$ through yet another first order transition. This proposition is further supported by noise data in $\chi'$, shown in the lower panels of Fig.2. Fluctuations in $\chi'$ signal for the ZFC state are measured using a lock-in amplifier with a flat band filter. The noise signal increases abruptly at T$_{pl}$, remains large in the PE region and finally decreases above T$_p$ with an associated vanishing of metastability and history dependence in $\chi'$ (all data pertaining to this not shown here). The present experiment primarily probes the response of the pinned FLL ; thus the fluctuating signal in $\chi'$ corresponds to transitions among the metastable (pinned) states of FLL accessible from the ZFC branch for a given h$_{ac}$. [Note that the ZFC branch itself is a metastable state with a deep minimum, and a large barrier separates it from, say, the FC state]. The flux flow noise studies [11,24] on moving FLL have shown that, slightly above the depinning current, the PE region in 2H-NbSe$_2$ is especially noisy due to spatially inhomogeneous plastic flow channels that, however, persist in the pinned state. Thus, the abrupt increase in the noise signal at T$_{pl}$ would imply the sudden enhancement of transitions among metastable states associated with the slow dynamics of the dislocations/grain boundaries in the fractured flux line lattice. The noise at T < T$_{pl}$ is very small due presumably to the paucity of dislocations below the onset of fracturing / melting of FLL, while the collapse of the noise at higher T is consistent with a phase cancellation of a large number of incoherent fluctuations in the fully disordered state.

A detailed study of path dependence in $\chi'$ has yielded striking results. Figs. 3(a) and 3(b) show $\chi'$ with the same h$_{ac}$, but following different temperature scan plans involving both ZFC and FC procedures. [The dotted lines ( with data points omitted ) in Fig.3(a) and Fig.3(b) identify the $\chi'$ curves recorded while warming up from the lowest T in the ZFC state; these data are also contained in Figs. 2(a) and 2(b)]. A FLL prepared in ZFC state at the lowest T was warmed up to a preselected temperature (T$_I$ or T$_{II}$ or T$_{III}$) and $\chi'$ was then recorded while cooling down (FCC) from the above mentioned temperatures. These temperatures were selected on a ZFC $\chi'$ curve, such that, (i) T$_I$ < T$_{pl}$, (ii) T$_{pl}$ < T$_{II}$ < T$_p$ and (iii) T$_p$ < T$_{III}$ < T$_c$. The following features are noteworthy in Figs. 3(a) and 3(b):
(a) For T$_I$ < T$_{pl}$ , the $\chi'$ data during the cool down cycle retraces the response during warm up (ZFC) cycle.
(b) For T$_p$ < T$_{III}$ < T$_c$, the $\chi'$ data (see triangles) during the cool down cycle (from T$_{III}$) initially retraces (down

to $T_p$) the ZFC $\chi'$ curve, but eventually at $T < T_{pl}$, the cool down $\chi'$ curve resembles that recorded while warming up in the FC mode ( compare FCC curves in Figs.3(a) and 3(b) with FCW $\chi'$ curves in Figs. 2(a) and 2(b)). In the intervening temperature region, $T_{pl} < T < T_p$, $\chi'$ ( in Fig. 3(a) / Fig.3(b) ) is different from both the ZFC and FCW curves in the corresponding upper panels of Fig.2. In fact, $\chi'$ in the intermediate region is more diamagnetic (i.e., larger $J_c$ and concomitantly smaller $V_c$) than even that recorded in FCW mode. That the cool down response is retraced ( for $T_p<T<T_{III}$ ) along the warm up cycle once again demonstrates that FLL system is disordered in equilibrium above $T_p$ and the supercooling of disordered phase does occur (on traversing back from $T_{III}$) across $T_p$.

(c) The most interesting phenomenon is observed when the FLL is cooled from $T_{II}$, i.e., from the partially fractured but not wholly disordered state. $\chi'$ starts to increase (see open circles in Figs. 3(a) and 3(b)) on lowering T (from $T_{II}$), it overshoots the ZFC cycle, in an attempt to recover the more ordered FLL as in the ZFC branch. However, at a temperature near, but, slightly above $T_{pl}$, it drops precipitously to a value close to the FCW branch instead (compare data in Figs. 3(a) and 3(b) with curves in Figs. 2(a) and 2(b)). This striking result suggests the following scenario. With increasing T towards the PE region, the FLL softens, the energy needed to create dislocations decreases and the lattice spontaneously fractures at $T_{pl}$. Upon lowering T from within the PE region, the lattice stiffens and stresses build up. The system fails to drive out the dislocations in order to heal back to the ZFC state. Instead, it fractures further in order to relieve stresses and reaches the other available metastable state, namely the FC state. This yields an open hysteresis curve in T (i.e., one cannot recover to the original ZFC state by cycling in T), which is highly unusual and not seen in typical first order transitions in the absence of disorder. Whether differences in entropic contributions to the free energies of the FC and ZFC states are responsible for such a process is unclear.

Combining results of dc and ac measurements, we plot the magnetic phase diagrams in Figs.4(a) and 4(b). At low H and T, a nearly defect-free lattice exists, akin to an elastic or Bragg glass [25]. At ($H_{pl}$,$T_{pl}$) line, the FLL system fractures into a plastically deformed solid [9] that we shall call a plastic glass likely similar to the Gingras-Huse [26] vortex glass. At even a higher line given by ($H_p$,$T_p$) values, the FLL further fractures into a *fully disordered* state with little difference between the ZFC and FC states. This condition is still *not fully reversible* and may thus be called a pinned flux line liquid or a pinned amorphous state, the exact nature of which is uncertain. Finally, above the irreversibility ($H_{irr}$,$T_{irr}$) line, the FLL system reaches an unpinned liquid state.

A variety of other experiments have also been performed for both systems. In one experiment, the FLL in the FC mode was subjected to an $h_{ac}$ of about ten times the value with which $\chi'$ data in Figs. 1 to 3 were recorded. This perturbation changes $\chi'$ to a value nearly the same as that observed in the ZFC state. Upon further increase of T, $\chi'$ follows the ZFC curve. In another experiment, for a FLL in the ZFC state at $T < T_{pl}$, a high current was pulsed through the sample heater such that the sample temperature momentarily cycles through a value lying above $T_{pl}$. This act is sufficient to bring the FLL system from the ZFC state back to the FC state. These procedures constitute a "switching" of the FLL between weakly and strongly irreversible states and will be described in detail elsewhere.

To conclude, the results in both systems appear consistent with the loss of order of the FLL with increasing thermal fluctuations, softening of the lattice and the resulting dominance of pinning. The observation of a first order transition marking the onset of the PE is proposed to be a result of a disorder-induced fracturing transition of the lattice. However, our magnetisation studies neither address the question whether GFFLO state exists in CeRu$_2$ nor elucidate the nature of its pinning centres.

Acknowledgement: We thank B. K. Chakrabarti, D. S. Fisher, D. A. Huse, Satya Majumder, N. Trivedi and V. M. Vinokur for useful discussions.
* Corresponding authors

FIG. 1. The panels (a) and (b) show M-H curves in $CeRu_2$ and 2H-$NbSe_2$ respectively at the temperatures indicated. In the latter crystal , the complete M-H cycle is shown, whereas in the former, only the anomalous opening and closing of the hysteresis bubble in the PE region is depicted. The insets in Figs. 1(a) and 1(b) show the in-phase ac ( f=211 Hz, $h_{ac}$ = 0.5 Oe (r.m.s)) susceptibility ($\chi'$) vs H.

FIG. 2. $\chi'(T)$ in crystals of $CeRu_2$ and 2H-$NbSe_2$ at dc fields as indicated. The upper panels (a) and (b) show $\chi'$ response recorded during warm up (W) from the lowest temperature ($\sim$ 4K) for FLL prepared in ZFC and FC modes. The lower panels (c) and (d) show the noise signal in $\chi'$ for FLL prepared in ZFC mode. The noise signal registers significant enhancements at $T_{pl}$ and $T_p$ and returns to the background level between $T_p$ and $T_c$.

FIG. 3. $\chi'(T)$ curves following different T scans. $T_I$, $T_{II}$ and $T_{III}$ identify the temperatures upto which a given sample was warmed up each time after preparing FLL in ZFC state at 4.2 K. The $\chi'$ data were recorded while cooling down (FCC) from $T_I$, $T_{II}$ and $T_{III}$ . The dotted lines in Figs. 3(a) and 3(b) sketch (respectively) the ZFC curves of Figs. 2(a) and 2(b).

FIG. 4. Magnetic phase diagrams in $CeRu_2$ and 2H-$NbSe_2$ depicting $H_{pl}$, $H_p$, $H_{irr}$ and $H_{c2}$ lines.